\documentclass[%
 reprint,
 amsmath,amssymb,
 aps,
prab,
floatfix,
]{revtex4-2}
\usepackage{amsmath}
\usepackage{amssymb} 
\usepackage{graphicx}
\usepackage{array}  
\usepackage{dcolumn}
\usepackage{bm}
\usepackage{booktabs}
\usepackage{multirow,ragged2e}
\usepackage{stfloats}
\usepackage{caption,subcaption}
\usepackage[dvipsnames]{xcolor} 
\usepackage{hyperref}
\usepackage{siunitx}
\begin{document}

\title{Development of Ultra High Power Compact X-Band Pulse Compressor}

\author{A. Dhar}
\email{adhar@slac.stanford.edu}
\author{M. A. K. Othman}
\author{V. A. Dolgashev}
\affiliation{SLAC National Accelerator Laboratory, Menlo Park,California,USA}

\date{\today}
\begin{abstract}
We demonstrate a new \SI{11.424}{\giga\hertz} SLED-type RF pulse compressor for powering high-gradient X-band photoinjectors with pulse lengths around \SI{20}{\nano\second}. RF pulse compression provides a practical path to higher peak power at the cost of pulse length for various applications such as RF deflectors for electron beam diagnostics on free electron lasers. Our new compact pulse compressor uses spherical cavities supporting axially symmetric TE modes which have minimal electric fields on the cavity surfaces, intended to improve high-power robustness as compared to existing compact spherical SLEDs which use a TE dipole mode. We present the design of this pulse compressor composed of two spherical cavities and a waveguide hybrid. The two cavities and hybrid have TE01 circular waveguide ports. This pulse compressor was built and high power tested at SLAC. These tests demonstrated a peak power of \SI{317}{\mega\watt} by compressing \SI{52}{\mega\watt}, \SI{1}{\micro\second} pulses generated by a SLAC XL-4 klystron representing, to our knowledge, the highest peak power achieved from a single-stage compact SLED RF pulse compressor at X-band. The full-width at half-maximum of this compressed pulse was \SI{27.1\pm1.1}{\nano\second}. We conjecture that this development demonstrates a viable route to reaching the high-gradient, short pulse regime for accelerating structures and RF photoinjectors.
\end{abstract}

\maketitle

\section{Introduction}
The performance of modern free-electron lasers such as the SLAC Linac Coherent
Light Source (LCLS) is driven by the brightness of the electron beam, which
improves with reduction in beam transverse emittance. Among the approaches for reducing the transverse emittance is increasing the
electric field on the cathode surface at the time of emission to mitigate
space-charge forces through rapid acceleration~\cite{Carlsten}. A major obstacle
to increasing cathode fields is RF breakdown, which disrupts RF pulses and
damages the cavity. The probability of RF breakdown reduces with decreasing
pulse length~\cite{design}. Recently, the Argonne Wakefield Accelerator
successfully achieved this increased electric field in a photoinjector by using
ultrashort ($<$\SI{20}{\nano\second}) RF pulses with high peak power to generate
cathode surface fields up to
\SI{400}{\mega\volt/\metre}~\cite{tan_demonstration_2022,ChenGun}.

In order to create such short-pulse electric fields in RF cavities, they need to
be powered by multi-MW RF sources with very short pulse lengths
($<$\SI{20}{\nano\second})~\cite{shortPulse}. Because klystrons capable of
supplying such short, multi-MW pulses are not available, RF pulse compression
provides a practical path to higher peak power at the cost of pulse length. One
type of RF pulse compressor, the so-called SLED (SLAC Energy Development), was
invented at SLAC~\cite{SLED}. SLED-type RF pulse compressors use low loss
cavities fed by external waveguides to store energy. This energy is then rapidly
discharged from the cavities, resulting in an increased peak power. The gain of
the SLED is determined by the external quality factor of the low loss cavities,
and the so-called charging time. During this charging time, the cavity is filled
with RF energy from the klystron. This stored energy is then compressed and
rapidly discharged by flipping the phase of the incoming RF pulse by 180
degrees. This phase flip takes finite time ($t_{\mathrm{flip}}$), which is
limited by the overall bandwidth of the RF system, and impacts the compression
gain of the SLED.

Over the years, SLAC has created numerous pulse compression systems for high
power RF applications, such as linear colliders~\cite{tantawi}. More recently,
SLAC built compact \SI{11.424}{\giga\hertz} SLEDs to power X-band transverse
deflectors for LCLS and LCLS-II~\cite{frazi,wang,Dolgashev:2021jcv}. Each of
these compact SLEDs use an over-moded spherical cavity operating with dipole
TE$_{114}$ modes. Other developments utilizing compact spherical cavities with
dipole modes are two-stage pulse compressors. These pulse compressors have
demonstrated peak powers up to \SI{320}{\mega\watt} at X-band and
\SI{1.2}{\giga\watt} at S-band~\cite{twoStageX,twoStageS}. However, the dipole
modes used in all of these spherical cavities have electric fields on the cavity
walls which can cause RF breakdowns, potentially limiting the output power of
these systems.

In order to increase peak output power while mitigating the risk of breakdown
associated with high cavity surface electric fields, we have developed a new
\SI{11.424}{\giga\hertz} SLED. This SLED uses two over-moded spherical cavities
with an axially-symmetric TE$_{023}$ spherical cavity mode. The chosen mode
minimizes surface electric field in the main cavity body, reducing breakdown
risk as compared to the dipole mode of afore-mention spherical cavities. The
compression gain of this SLED is determined by the ratio between the internal
and external quality factor of the cavities, known as the coupling factor
($\beta$). We chose the coupling factor of these cavities to be 25, allowing it to provide up to 6.5x compression gain when powered by
\SI{50}{\mega\watt} \SI{1}{\micro\second} pulses generated by a SLAC XL-4
klystron with a bandwidth of about \SI{100}{\mega\hertz}~\cite{XL4}.

The two spherical cavities of the SLED are connected through a previously
designed waveguide hybrid with circular ports~\cite{hybrid,modeConverter}. These
ports operate with the TE$_{01}$ circular waveguide mode. We have designed,
built and high power tested this new SLED at SLAC, demonstrating peak RF power
up to \SI{317}{\mega\watt}. While higher peak powers have been reported
using multi-stage pulse compression (up to \SI{320}{\mega\watt} at X-band with a
two-stage system~\cite{twoStageX} and \SI{1.2}{\giga\watt} at
S-band~\cite{twoStageS}), to our knowledge this is the highest peak power
achieved by a single-stage X-band SLED pulse compressor, and is realized in a
substantially more compact footprint than multi-stage systems.

\begin{figure}[h]
\centering
\includegraphics[width=\columnwidth]{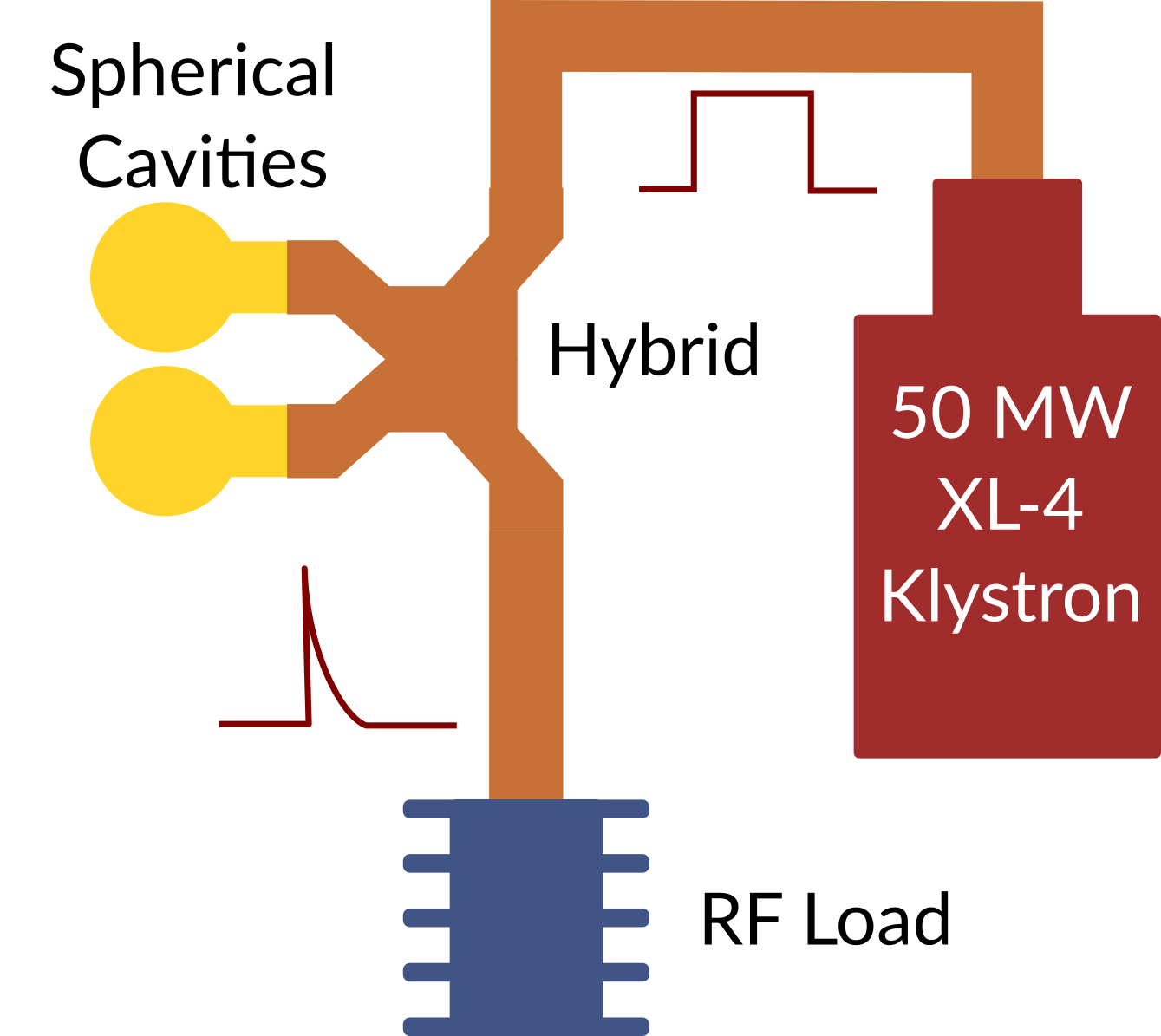}
\caption{Diagram of the experimental apparatus under test at SLAC. Power from
the SLAC XL-4 \SI{50}{\mega\watt} klystron as square pulses is directed towards
two spherical cavities connected through a waveguide hybrid. Power stored in
these cavities is then compressed into short intense spike-like via a flip in RF
phase and is sent into a dry RF load for dissipation.}
\label{fig:diagram}
\end{figure}
\section{Design of the TE$_{023}$ Pulse Compressor}

\subsection{Cavity Mode Selection and Geometry}
In the design of high-power pulse compression cavities, there are two crucial
criteria to consider. These criteria are: (i) a very high internal quality
factor, and (ii) sufficient spacing between the operating mode and neighboring
parasitic modes that couple to the TE$_{01}$ circular waveguide mode. One way to
achieve these requirements is to use an overmoded spherical cavity. In high RF
power applications, it is beneficial to use an axially (cylindrically) symmetric
TE mode due to its low loss and minimal electric field on the cavity
surface. For our purposes, we used the TE$_{023}$ spherical mode, which has
rotational symmetry about the Z-axis to couple well to the TE$_{01}$ circular
waveguide mode which has the same symmetry. The radius of the spherical cavity
was then chosen to center this mode on \SI{11.424}{\giga\hertz}. The external
coupling into the cavity was set by an iris between a circular waveguide and the
cavity. In this design, the circular waveguide diameter is 1.5~inches
(\SI{38.1}{\milli\meter}), and the spherical cavity has an inner diameter of
4.037~inches (\SI{102.54}{\milli\meter}).

\subsection{High-Power RF Simulations}
\begin{figure}[htbp]
\centering
\begin{subfigure}[h]{\columnwidth}
\caption{}
\includegraphics[clip=true,trim={0 0 0 7},width=\textwidth]{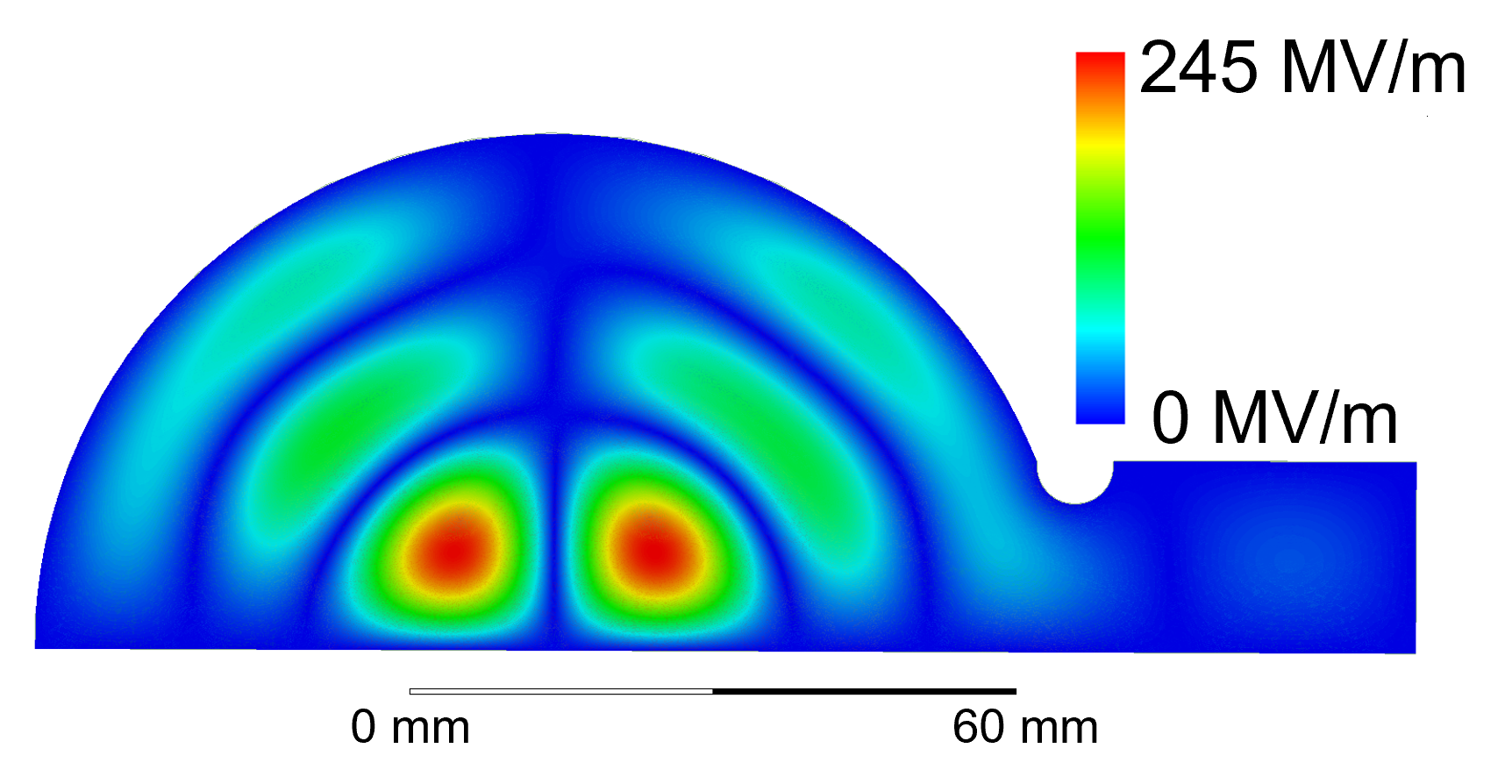}
\end{subfigure}
\begin{subfigure}[h]{\columnwidth}
\caption{}
\includegraphics[clip=true,trim={0 0 0 7},width=\textwidth]{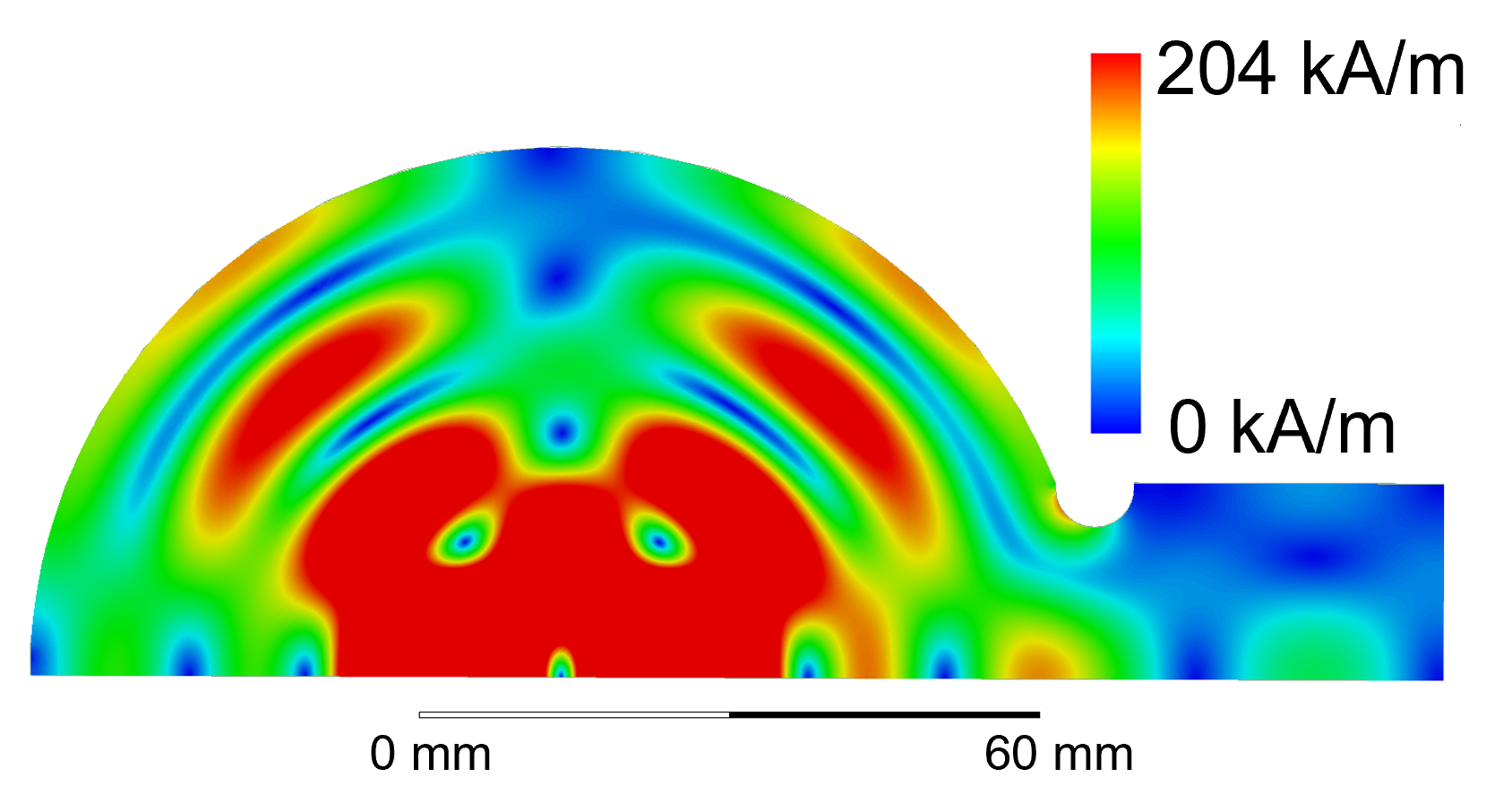}
\end{subfigure}\\
\begin{subfigure}[h]{\columnwidth}
\caption{}
\includegraphics[clip=true,trim={0 0 0 7},width=\textwidth]{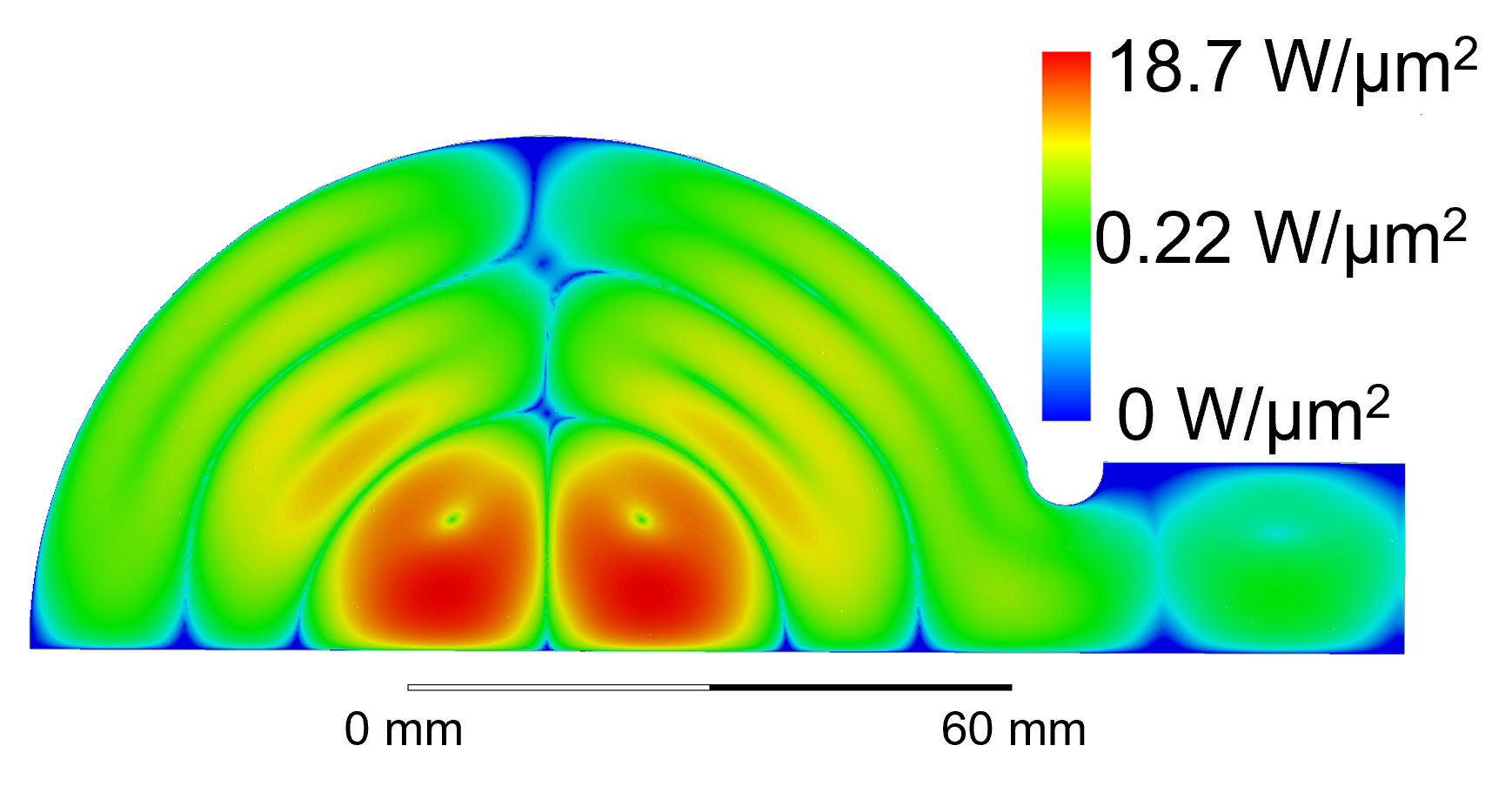}
\end{subfigure}
\begin{subfigure}[h]{\columnwidth}
\caption{}
\includegraphics[clip=true,trim={0 0 0 7},width=\textwidth]{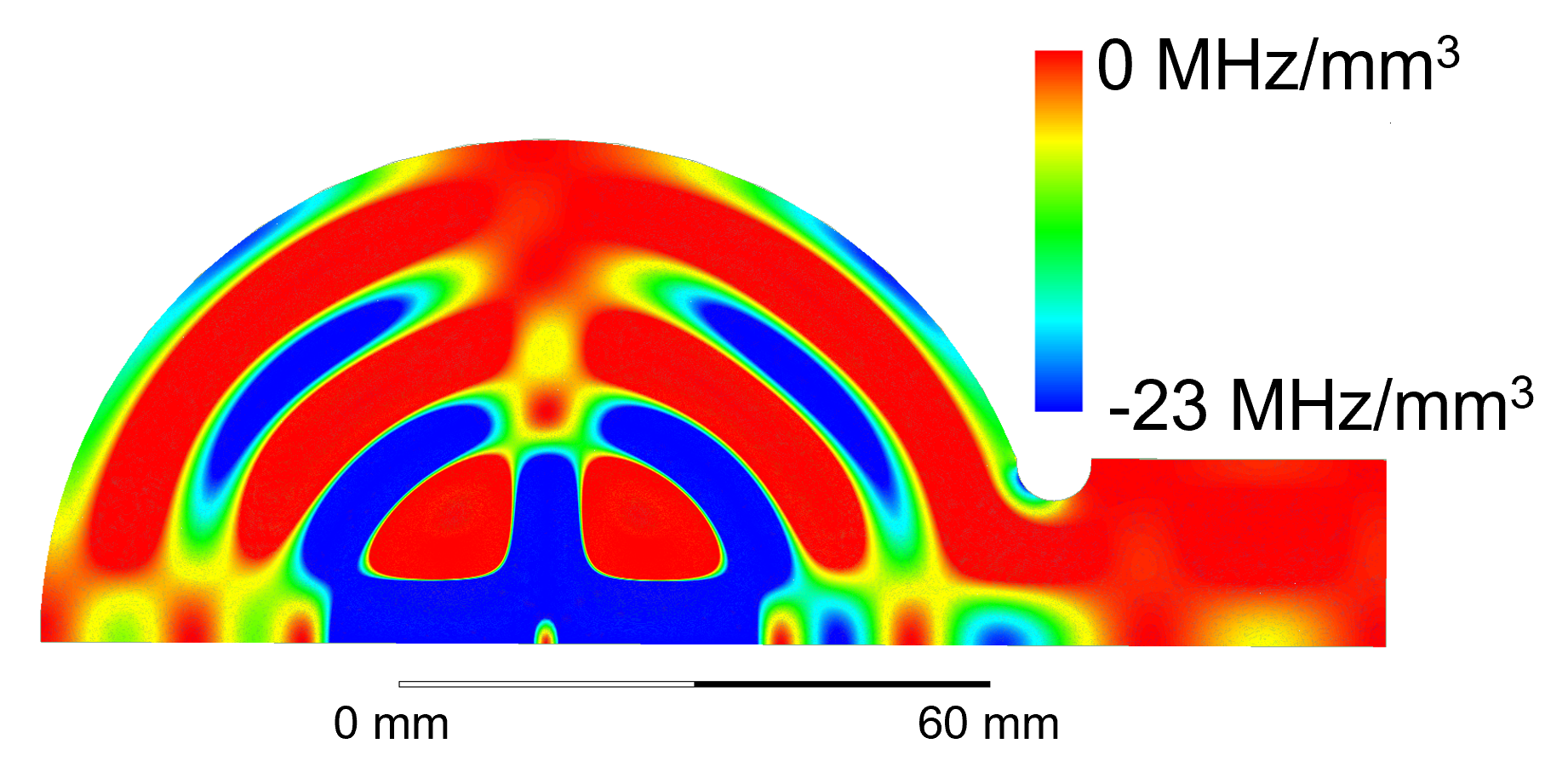}
\end{subfigure}
\caption{(a) Volume RF electric fields, (b) volume RF magnetic fields, (c) volume
modified Poynting vector, and (d) tuning sensitivity for a spherical cavity. A
symmetric half of the cavity is shown. Fields are normalized to input power of
\SI{50}{\mega\watt} and a coupling factor of $\beta=25$. The TE$_{023}$ mode
within the cavity is excited at \SI{11.424}{\giga\hertz}.
The electric field and modified Poynting vector are 0 on the cavity surface.}
\label{fig:HFSS}
\end{figure}

The high power RF performance of this spherical cavity was simulated using Ansys
HFSS~\cite{ANSYS}. To evaluate the performance, we assumed an input RF power of
\SI{50}{\mega\watt}. These simulated fields (shown in Figure~\ref{fig:HFSS})
were used to evaluate the risk of vacuum RF breakdowns and material damage. We
present high power performance simulation of the optimized SLED spherical
cavities to show peak surface electric ($E$) field, peak surface magnetic ($H$)
field, peak pulse surface heating, and a modified form of Poynting
vector~\cite{Nantista:2004qd,Sjobak:2014nta,grudiev2009new}.

The peak surface $E$ field on the cavity surface was calculated to compare with
the expected breakdown criterion of \SI{250}{\mega\volt/\meter}. This cavity
mode has minimal electric field on the surface, so this is not a major concern.
The peak surface $H$ field was calculated to determine the estimated pulsed
surface heating. This calculation was based on analysis that relates peak
surface $H$ field and pulse width to a temperature rise for a square RF pulse
along a metal surface~\cite{pulsed}. We use this approximation knowing that our
input pulse is not square. The equation for \SI{11.424}{\giga\hertz} is given as
\begin{align}
\Delta T_{\mathrm{pulse}}=430|H_{\mathrm{peak}}|^2 \sqrt{t},\label{eq:pulseheat}
\end{align}
where $\Delta T_{\mathrm{pulse}}$ is in \SI{}{\celsius}, $H_{\mathrm{peak}}$ is
in \SI{}{\mega\ampere/\meter}, and $t$ is in \SI{}{\micro\second}. The peak
surface magnetic fields induce surface heating of
\SI{25}{\celsius}, while the threshold for surface
damage is \SI{50}{\celsius}~\cite{Laurent}. This surface heating calculation is
an overestimation given that our RF pulse in the cavity is not square.

As a further measurement of power flow, the Poynting vector magnitude along the
surfaces was also evaluated. However, this magnitude was calculated in a
specific fashion to compare with the empirical breakdown analysis of existing
high-gradient accelerating structures. This modified Poynting vector was
calculated as $S_c = |\mathrm{Re}(S)|+0.2 |\mathrm{Im}(S)|$, which was then
evaluated against an established criterion for RF
breakdowns~\cite{grudiev2009new}. This criterion is based on the established
value of \SI{5}{\watt/\micro\meter^2}
for a \SI{200}{\nano\second} pulse. Further, using a scaling law developed for
traveling wave structures, the criterion is
\SI{2.3}{\watt/\micro\meter^2} for
a pulse length of \SI{2}{\micro\second}.
Finally, we calculated the sensitivity of cavity frequency $f$ to mechanical
deformations using cavity perturbation theory~\cite{Slater}.
This can be determined from the following equation
\begin{align}
\frac{\Delta f}{f}\approx\frac{\iiint_{\Delta V}\mu |H_0|^2 -
\epsilon|E_0|^2dV}{U_0},\label{eq:tuning}
\end{align}
where the integral of the difference in magnetic and electrical energy is taken
over the relative perturbation in volume.

The TE023 mode in this cavity has an internal quality factor ($Q_0$) of 82,000.
The coupler provides a coupling factor of 25 and loaded quality factor ($Q_l$) of
3380. Based on these parameters, the effective compression gain was evaluated
for a range of charging times, determining that the maximum compression gain
possible for this SLED should be between 6.5x to 6.8x depending on charging time
ranging from \SI{300}{\nano\second} to \SI{1500}{\nano\second}. The design
parameters for this cavity and SLED design are summarized in
Table~\ref{tab:SLED}.

\begin{table}[h]
\centering
\caption{Design parameters of RF pulse compressor operating at
\SI{11.424}{\giga\hertz}.}
\begin{tabular}{ccc}
\toprule
\multicolumn{3}{c}{\textbf{Cavity Parameters}} \\
\midrule
$Q_0$ & $Q_l$ & $\beta$ \\
82000 & 3380  & 25      \\
\midrule
\multicolumn{3}{c}{\textbf{SLED Parameters}} \\
\midrule
Pulse length & Charge time & $t_{\mathrm{flip}}$ \\
\SI{1}{\micro\second} & $>$\SI{250}{\nano\second} & \SI{10}{\nano\second} \\
\midrule
\multicolumn{3}{c}{\textbf{High Power RF Parameters at \SI{50}{\mega\watt}}} \\
\midrule
$H_{\mathrm{peak}}$ & $S_c$ & $\Delta T_{\mathrm{pulse}}$ \\
\SI{204}{\kilo\ampere/\meter}
  & \SI{0.22}{\watt/\micro\meter^2}
  & \SI{25}{\celsius} \\
\bottomrule
\end{tabular}
\label{tab:SLED}
\end{table}

\section{Fabrication and Cold-Test Characterization}

\subsection{Cavity Tuning}
The SLED cavities were fabricated out of OFE Copper and brazed. After
fabrication, each TE$_{023}$ spherical cavity was tuned to ensure optimal RF
pulse compression at \SI{11.424}{\giga\hertz}. During the tuning process the
cavity was filled with nitrogen gas and its temperature was monitored with
thermocouples. The specific target frequency is determined by accounting for the
relative change in frequency due to temperature and pressure of dry nitrogen. For
a final operating target of \SI{11.424}{\giga\hertz} in vacuum at
\SI{15}{\celsius}, and a measurement temperature of \SI{20}{\celsius}, our target
frequency for tuning was \SI{11.420}{\giga\hertz}. During the experiment, the
temperature of each cavity was independently controlled by dedicated water
chillers.


\begin{figure}[h]
\centering
\includegraphics[width=\columnwidth]{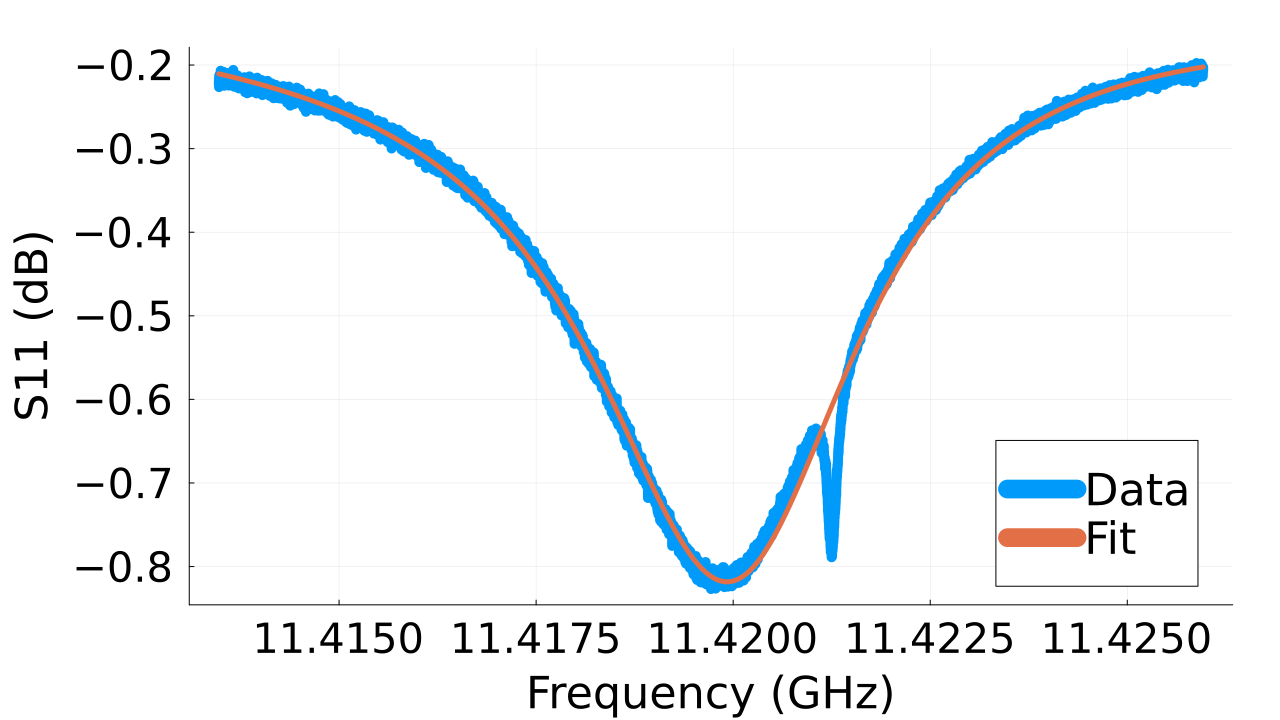}
\caption{Measurement of reflection coefficient for a spherical cavity after tuning. The cavity was tuned to \SI{11.4201}{\giga\hertz} at \SI{20}{\celsius} filled with dry nitrogen. Based on a fit to the data (orange), the internal quality factor was determined to be 71854, with a coupling factor of 25.8. The internal quality factor is lower than the expected design value, but with the right coupling factor this would still be sufficient for pulse compression. Measurements after tuning (blue) show a small parasitic mode near the
target frequency of \SI{11.42}{\giga\hertz}. The parasitic mode seen here couples to the TE$_{21}$ circular waveguide mode.}
\label{fig:tuned}
\end{figure}

\subsection{Parasitic-Mode Identification}

\begin{figure}[htbp]
\centering
\includegraphics[width=\columnwidth]{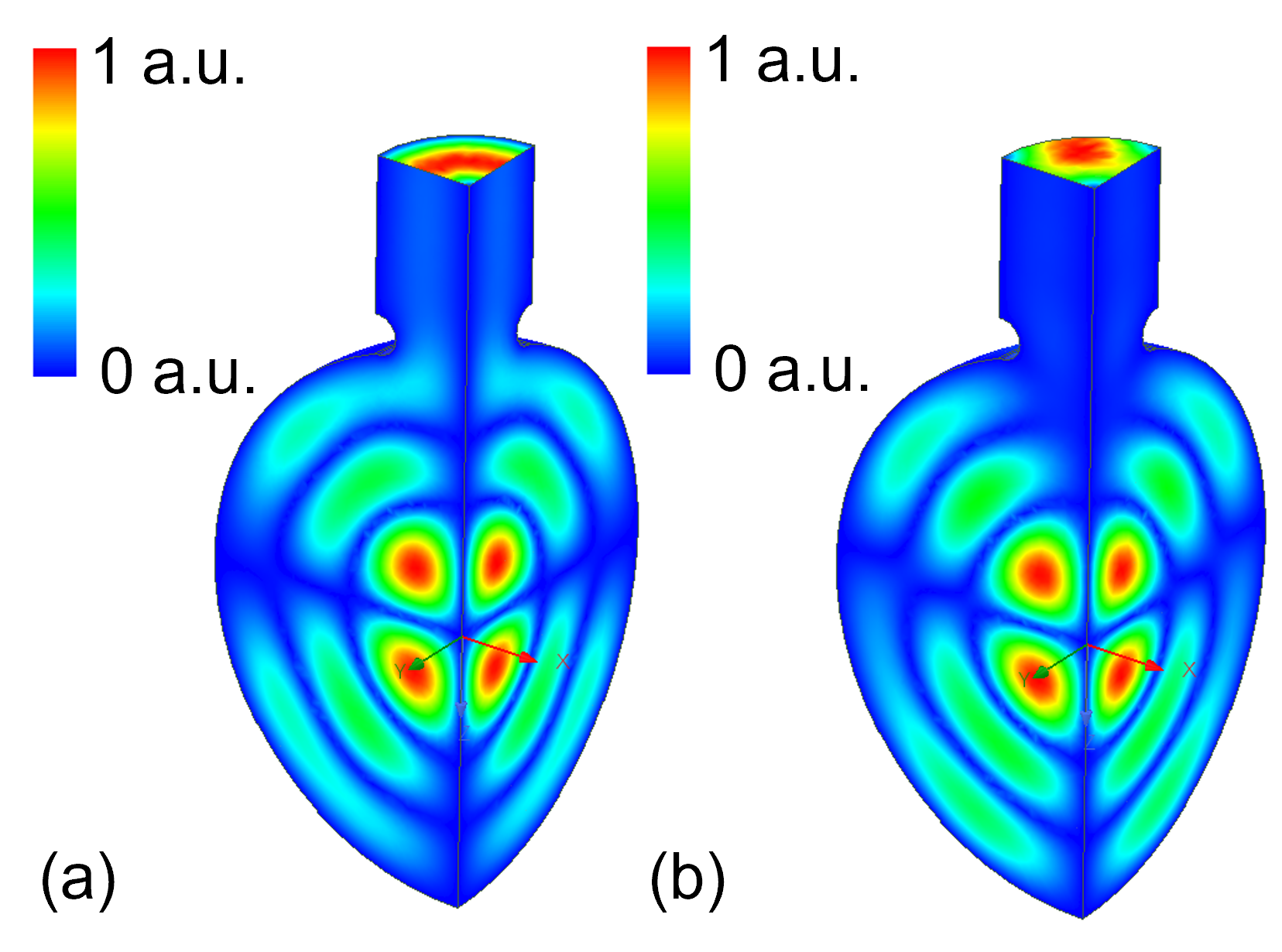}
\caption{HFSS-simulated electric field magnitude distributions comparing (a) the
operating TE$_{023}$ cavity mode and (b) the nearby parasitic mode. In the
circular waveguide port (top), the operating mode exhibits the azimuthally
symmetric field pattern characteristic of the TE$_{01}$ circular waveguide mode,
whereas the parasitic mode shows an asymmetric field pattern characteristic of
the TE$_{21}$ quadrupole circular waveguide mode. Because the hybrid drives the
ports with high TE$_{01}$ mode purity, the TE$_{21}$-coupled parasitic mode is not
appreciably excited.}
\label{fig:modecompare}
\end{figure}

The results of this tuning are shown in Figure~\ref{fig:tuned}. After tuning both
cavities, a parasitic mode remained approximately \SI{5}{\mega\hertz} above
the operating mode, within the bandwidth of the klystron.
A fit to the measured reflection coefficient gives this parasitic mode a
loaded quality factor of 57380, compared to 3380 for the operating TE$_{023}$
mode. We identified through HFSS simulations that this parasitic mode
couples to the TE$_{21}$ quadrupole circular waveguide mode instead of the
TE$_{01}$ circular waveguide mode. Figure~\ref{fig:modecompare} compares the
simulated field distributions of the two modes: the operating mode presents an
azimuthally symmetric field in the circular waveguide port consistent with the
TE$_{01}$ mode, while the parasitic mode presents an asymmetric TE$_{21}$-like
field. Because the hybrid powering the two spherical cavities uses mode
conversion between the TE$_{01}$ circular and TE$_{10}$ rectangular waveguide
modes with high mode purity, this TE$_{21}$-coupled parasitic mode is not driven
appreciably. Therefore we do not expect this parasitic mode to couple strongly
during high power testing.

\section{High-Power Experiment}
\subsection{Experimental Setup}
\begin{figure}[htbp]
\centering
\begin{subfigure}[h]{\columnwidth}
\includegraphics[width=\textwidth]{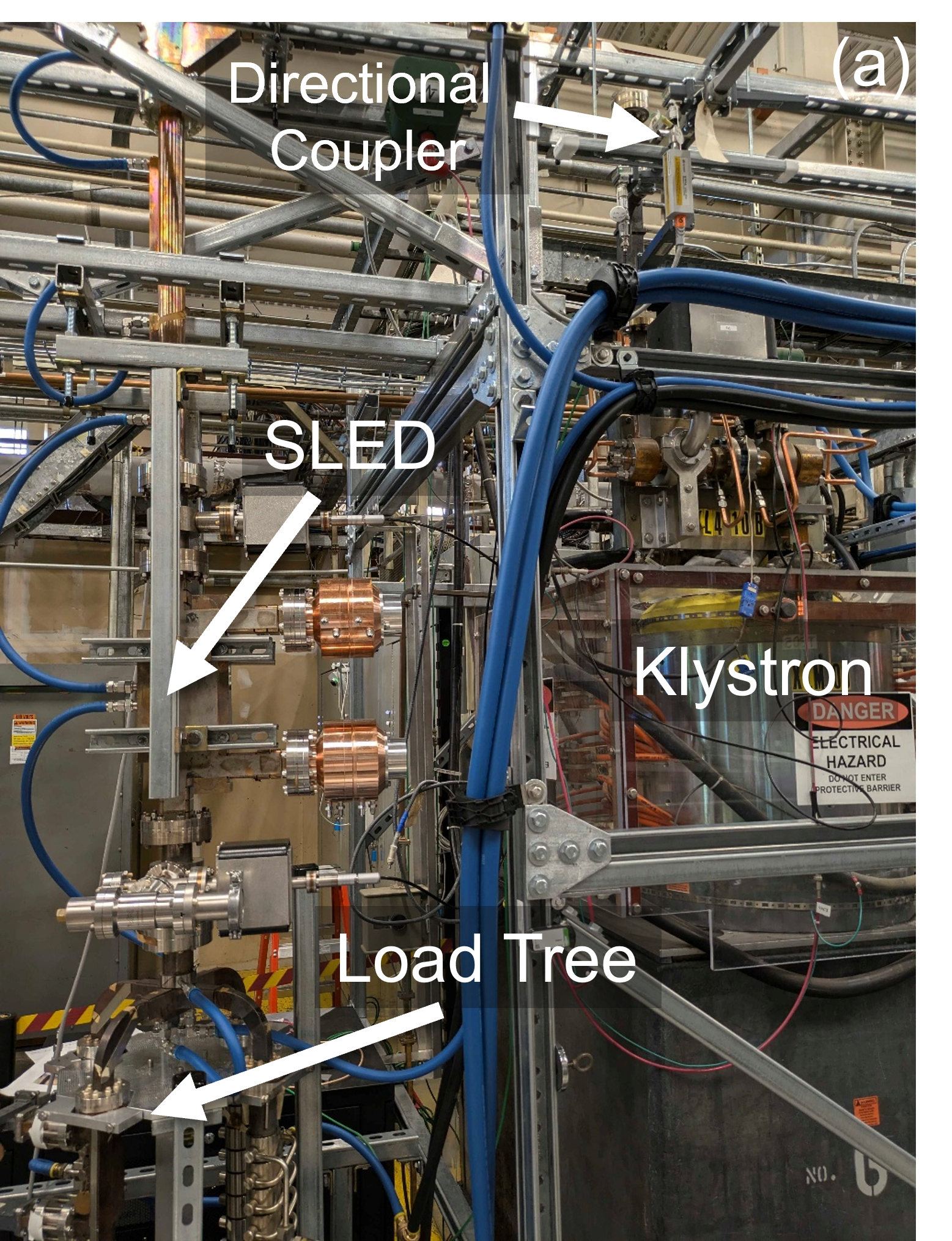}
\end{subfigure}
\begin{subfigure}[h]{\columnwidth}
\includegraphics[width=\textwidth]{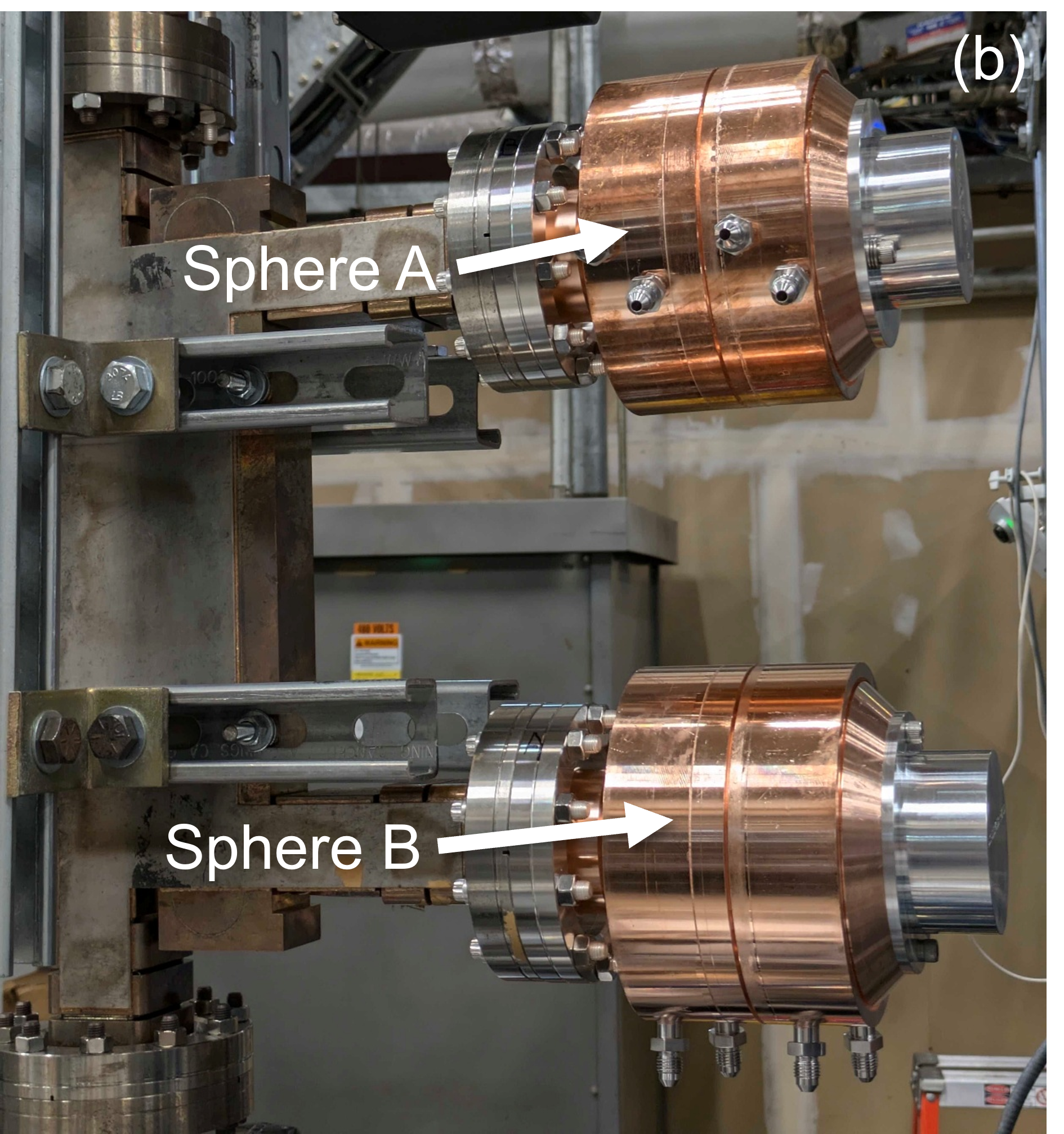}
\end{subfigure}
\caption{(a) Overview of experimental apparatus at SLAC. Power from the SLAC XL-4
\SI{50}{\mega\watt} klystron is directed to the SLED and then a so-called load
tree for dissipation. This load tree is a device which uses a four way splitter to
divide power between four RF dry loads. (b) Closeup of the two SLED spherical
cavities.}
\label{fig:SLED}
\end{figure}

Once tuned, the cavities were installed onto a waveguide hybrid with circular
ports to form the SLED system. The RF pulses from the spherical cavities after the
phase flip combine in the hybrid with the same amplitude and phase and the
combined power flows out of the hybrid. This SLED was then connected to the output
of a SLAC XL-4 \SI{50}{\mega\watt} klystron. An overview of the assembled high
power test setup is shown in Figure~\ref{fig:SLED}. The klystron output was
directed through vacuum WR90 waveguide, and down towards the SLED and RF load.
This load has a four way splitter to divide power between four stainless steel RF
dry loads. One of the load tree's four arms had a directional coupler, allowing
for measurements of the uncompressed power from the klystron and the compressed
power from the SLED.

All relevant RF components were temperature stabilized via a cooling loop of house
water, except for the cavities. Each cavity was connected to its own water
chiller, allowing them to be independently tuned via temperature. The flexibility
of this cooling circuit was vital to ensuring that both cavities remained tuned
during operation.

\subsection{Conditioning and RF Diagnostics}

In order to commission the pulse compression system, the SLED cavities were
thermally-detuned to perform initial conditioning at low power. This involved
slowly increasing the output power of the klystron from \SI{10}{\mega\watt} to \SI{50}{\mega\watt}, and pulse
length from \SI{100}{\nano\second} to \SI{1}{\micro\second}. During
this progression the vacuum was monitored for any sign of significant breakdowns
or arcing to allow the waveguide network to condition. As a result, the system
was able to operate with the klystron output power of \SI{50}{\mega\watt},
\SI{1}{\micro\second} pulse length, and \SI{120}{\hertz} repetition rate.

With the system conditioned, we characterized the compressed output using a
dedicated set of RF diagnostics. For the measurement of RF power, we used
calibrated peak power meters, fast digitizers to capture down-mixed RF signals,
and crystal detectors. The peak power meters have a time resolution of
\SI{10}{\nano\second}, while the digitizer has a resolution of
\SI{0.5}{\nano\second}. To tap off power from the high power waveguide circuit, we
have used two waveguide directional couplers: one after the klystron output, and
another in an arm of the load tree. Each signal was routed to both a peak power
meter and a down-mixer. All signals were then processed using LabView
software~\cite{labview2024}. While the power meters provided calibrated
measurements of forward power from the klystron, they did not have the temporal
resolution required to resolve the peak of the compressed pulse. Thus we have
calibrated the mixers against the power meters for our measurements of the peak
compressed power. The mixers were calibrated by detuning the SLED and ensuring
that a long RF pulse with a flat top measured out of the klystron and out of the
SLED were equal in terms of power level on both the peak power meters and the
digitizers.

\section{Results and Discussion}
\subsection{Peak Power and Compression Gain}
\begin{figure}[htb]
\centering
\centering
\begin{subfigure}[h]{\columnwidth}
\caption{}
\includegraphics[clip=true,trim={0 0 0 7},width=\textwidth]{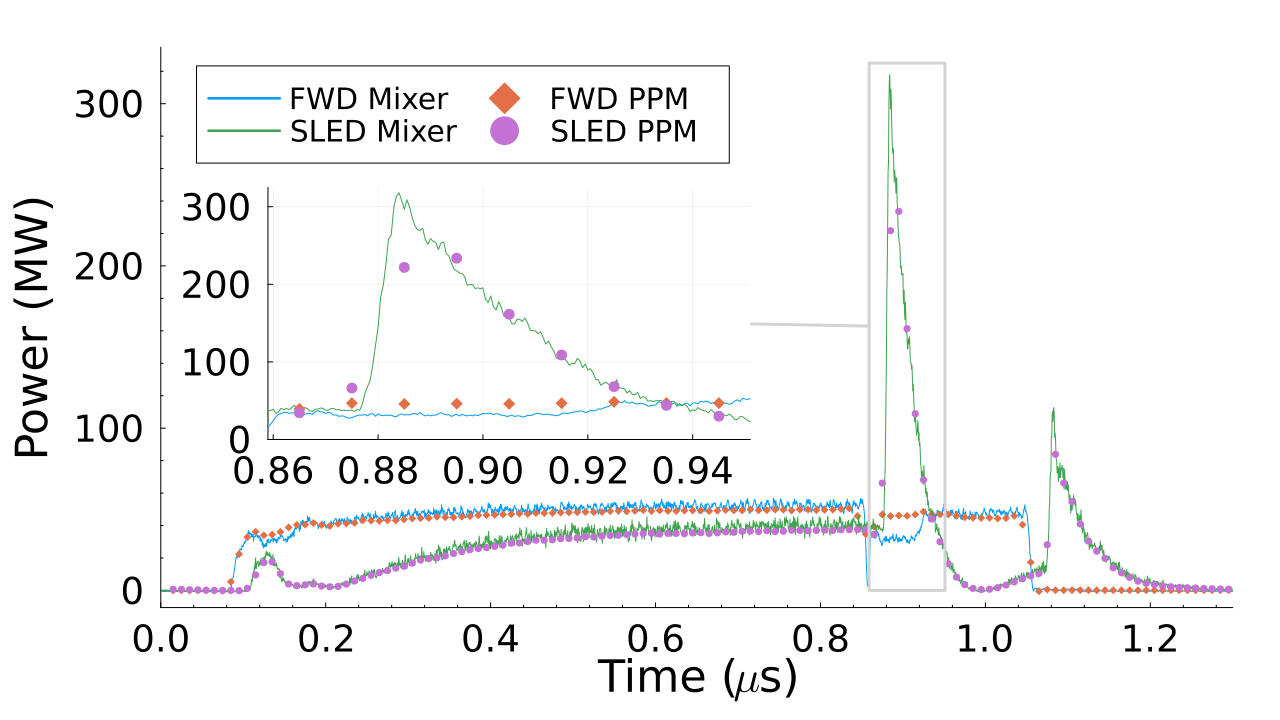}
\end{subfigure}
\begin{subfigure}[h]{\columnwidth}
\caption{}
\includegraphics[clip=true,trim={0 0 0 7},width=.9\textwidth]{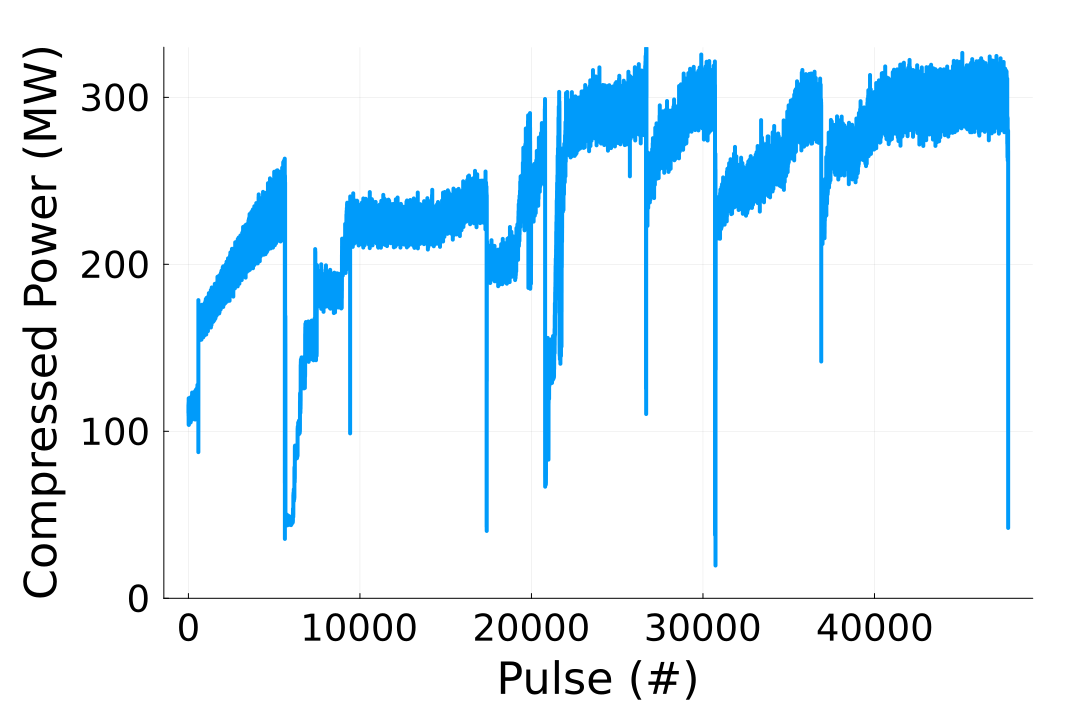}
\end{subfigure}
\caption{(a) RF measurements of output pulses from the klystron and the SLED. The mean forward klystron power (blue) just before the phase flip is \SI{52\pm2}{\mega\watt}, and the peak compressed power (green) is \SI{317\pm8}{\mega\watt}, yielding a compression factor of \SI{6.1\pm0.2}{}. Dots denote data from peak power meters, and lines represent data from down-mixed signals. (b) Progression of compressed power as measured by calibrated mixers over the last three days of conditioning, with operation reaching an output power over \SI{300}{\mega\watt}. Sharp drops in power were precipitated by vacuum interlock trips in other sections of the RF network, namely near the rectangular to circular waveguide mode converter out of the klystron.}
\label{fig:pulse}
\end{figure}

After initial conditioning and calibration, the SLED cavities were thermally-tuned
to resonance using chillers. At the resonance, the compressed power was maximized
and the reflected power back to the klystron was minimized. The system was
then fully conditioned over ten days to operate at \SI{52}{\mega\watt} output power over
\SI{1}{\micro\second} at \SI{120}{\hertz} rep rate. In order to compress the
klystron pulse, its phase needs to be flipped by 180 degrees after the SLED
cavities are filled. This phase flip was realized with an arbitrary waveform
generator and an RF mixer which modulates the RF pulse at the klystron input. We
monitored the phase flip via the down-mixed signals from the directional couplers,
which were digitized and converted into amplitude and phase. This allowed us to
confirm a phase flip of \SI{179.97\pm0.94}{\degree} over a flip time
\SI{9.1\pm1.5}{\nano\second} and a full-width at half-maximum (FWHM) of
\SI{27.1\pm1.1}{\nano\second} for the compressed power measured over 10000
pulses. An example of this amplitude and phase data is shown in
Figure~\ref{fig:pulse}.

The compression gain for each pulse was evaluated by dividing the peak compressed
power by the mean input forward power from the klystron immediately before the
phase flip, accounting for uncertainty in the calibrated mixer measurement. With
optimal tuning and phase-flip timing, we observed a compression factor of
\SI{6.1\pm0.2}, which agrees well with the simulation prediction of 6.4. We also
determined that the pulse compression efficiency, defined as the total pulse
energy out of the SLED divided by the total input pulse energy, to be
\SI{86.6\pm0.2}{\percent}, which agrees well with the theoretical expectation of \SI{89}{\percent}
considering the additional losses within the hybrid of \SI{2.2\pm0.2}{\percent} caused by
higher order modes and RF dissipation.
\subsection{Breakdown and Vacuum Behavior}

We did not directly measure a breakdown rate in this experiment. Instead, we
monitored vacuum pressure and reflected power as proxies for breakdown activity,
advancing conditioning whenever vacuum levels remained stable. The rate of
conditioning was primarily limited by vacuum activity in the waveguide mode
converter connecting the output WR90 waveguide from the klystron to the circular
waveguide feeding the SLED, rather than by activity in the SLED cavities
themselves. Any breakdown within the cavities would have produced a corresponding
loss of transmitted power registered by the directional coupler in the load tree;
we did not observe power loss consistent with breakdown originating in the
cavities. The sharp drops in compressed power shown in Figure~\ref{fig:pulse}(b)
were each precipitated by vacuum interlock trips localized to the mode converter
and other upstream sections of the RF network, and were not associated with the
SLED cavities. Following each trip, the affected section was reconditioned and full
compressed power was recovered. These observations are consistent with the low
surface electric field of the TE$_{023}$ mode, which vanishes on the cavity wall,
and support our expectation that the SLED cavities were not the
breakdown-limiting element of the system. They also confirm our prediction that the
parasitic mode observed in cold testing would not present any issues during high
power operation.

\section{Conclusion}
We have built and experimentally demonstrated the performance of an ultra high peak
power \SI{11.424}{\giga\hertz} SLED-type RF pulse compressor. This SLED has
achieved peak compression gain of 6.1 over \SI{27.1\pm1.1}{\nano\second}, and has
generated peak output power up to \SI{317\pm8}{\mega\watt}, the highest peak
power from a single-stage X-band SLED pulse compressor to our knowledge, and in a
substantially smaller physical footprint than the two-stage pulse compressor
that reached comparable power at the same frequency~\cite{twoStageX}. We
conjecture that this development provides a viable route to reaching the
high-gradient, short pulse regime for accelerating structures and RF
photoinjectors~\cite{Tan2025}. Increasing the cathode gradient on photoinjectors
has the potential to improve electron beam brightness which translates to improved
XFEL performance.

\section{ACKNOWLEDGMENTS}
The authors would like to thank Brad Shirley, Wei-Hou Tan, Adam Callman, John P
Eichner, and Emilio Nanni for their support. This work is supported by U.S.
Department of Energy Contract No. DE-AC02-76SF00515.

\bibliography{WEPS37}
\end{document}